\documentclass[lettersize,journal]{IEEEtran}
\usepackage{graphicx}
\usepackage{multirow}
\usepackage[dvipsnames]{xcolor}
\usepackage{color}
\usepackage[linesnumbered,ruled,vlined]{algorithm2e}
\usepackage{geometry}
\usepackage{xcolor,colortbl}
\usepackage{subcaption}
\usepackage{tikz}

\usepackage{pifont}
\usepackage{hyperref}
\usepackage[activate]{microtype}
\usepackage[normalem]{ulem}
\usepackage{authblk}
\usepackage{booktabs}

\usepackage{url}
\usepackage[dvipsnames]{xcolor}
\usepackage{soul}

\usepackage[separate-uncertainty = true,multi-part-units = repeat]{siunitx}
\usepackage{array,amsfonts,amsmath}
\usepackage{caption}
\ifCLASSOPTIONcompsoc
  \usepackage[nocompress]{cite}
\else
  \usepackage{cite}
\fi
\ifCLASSINFOpdf
\else
\fi
\begin{document}

\title{Non-Speech Emotions Detection using Edge Computing: Applications and Case Study}
\title{Non-Voiced Audio Emotions Detection using Edge Computing: Applications and Case Study}
\title{Beyond Words: Detecting Emotions in Non-Speech Audio With Edge Computing}
\title{Beyond Words: Non-Voiced Audio Emotion Recognition With Edge Computing}
\title{Emotions Beyond Words: Non-Speech Audio Emotion Recognition With Edge Computing}
\title{Going Beyond Words in Emotion Detection: Non-Speech Audio Emotion Recognition With Edge Computing}
\title{Emotion Detection Beyond Words: Non-Speech Audio Emotion Recognition With Edge Computing}
\title{Finding Emotions Beyond Words: Non-Speech Audio Emotion Recognition With Edge Computing}
\title{Emotions Beyond Words: Non-Speech Audio Emotion Recognition With Edge Computing}

\author[1]{Ibrahim Malik}
\author[2]{Siddique Latif\thanks{Email: siddique.latif@qut.edu.au}}
\author[3]{Sanaullah Manzoor}
\author[4]{Muhammad Usama}
\author[5]{Junaid Qadir}
\author[2]{Raja Jurdak}


\affil[1]{EmulationAI}
\affil[2]{Queensland University of Technology (QUT), Brisbane, Australia}
\affil[3]{University of the West Scotland, United Kingdom}
 \affil[4]{National University of Computer and Emerging Sciences (NUCES), Pakistan}
 \affil[5]{Qatar University, Doha}
 

\maketitle
\begin{abstract}

Non-speech emotion recognition has a wide range of applications including healthcare, crime control and rescue, and entertainment, to name a few. Providing these applications using edge computing has great potential, however, recent studies are focused on speech-emotion recognition using complex architectures. In this paper, a non-speech-based emotion recognition system is proposed, which can rely on edge computing to analyse emotions conveyed through non-speech expressions like screaming and crying. In particular, we explore knowledge distillation to design a computationally efficient system that can be deployed on edge devices with limited resources without degrading the performance significantly. We comprehensively evaluate our proposed framework using two publicly available datasets and highlight its effectiveness by comparing the results with the well-known MobileNet model. Our results demonstrate the feasibility and effectiveness of using edge computing for non-speech emotion detection, which can potentially improve applications that rely on emotion detection in communication networks. To the best of our knowledge, this is the first work on an edge-computing-based framework for detecting emotions in non-speech audio, offering promising directions for future research.

\end{abstract}


\IEEEpeerreviewmaketitle

\section{Introduction}
\label{sec:introduction}
\IEEEPARstart{T}{he} age of the Internet of Things (IoT) is upon us. The raging increase in IoT devices and the race among tech manufacturers to capture the market share has reached a point where the communication systems are struggling to fulfil the quality of service and experience requirements. 
The merging of Artificial Intelligence (AI) with the IoT has resulted in a plethora of practical applications in recent years. These applications span a wide range of fields, from image classification to stable diffusion \cite{croitoru2022diffusion} and speech recognition to real-time speech generation \cite{li2022recent, wali2022generative}. 
The healthcare industry has seen significant progress in disease detection, with AI outperforming human doctors in the early detection of disease \cite{hu2018deep}. Data collection and cleaning, as well as urban computing \cite{zheng2014urban}, have also benefited from this combination. Additionally, voice assistants and adaptive emotion recognition \cite{latif2022ai} are just a few of the many other applications that have emerged as a result of the fusion of AI and IoT. These developments have unprecedented levels of data storage and computational requirements. The traditional communication system design was not enough to fulfil the needs of these data and compute-hungry applications. It gave rise to cloud computing \cite{dillon2010cloud}, the backbone of AI-enabled IoT applications, and the widespread adoption of IoT applications is also credited to cloud computing technologies.


Edge computing is a distributed computing paradigm that decreases the data transmission load to the cloud by bringing enterprise applications near the data sources such as IoT devices or edge servers. This proximity to data at its sources has the potential to bring strong business benefits including better response times, improved bandwidth availability, faster decision-making, and privacy preservation. The development of computational technologies like graphics processing units, tensor processing units, etc., makes it feasible to offload some computational tasks to potent edge servers. When it comes to real-time services/applications (e.g., traffic monitoring systems, facial recognition, control system applications), latency, quality of service, and experience become increasingly critical. In particular, low latency is crucial in emotion recognition applications, where the computing device needs to classify the user's emotional state from given input audio or visual data for a particular application. Real-time emotion identification becomes even more important in a life-threatening serious situation. In such cases, edge computing has the potential to meet the latency requirements. In this work, we present an edge computing-based non-speech emotion detection system. 

\begin{figure*}[!ht]
\centering
\includegraphics[width=.65\textwidth]{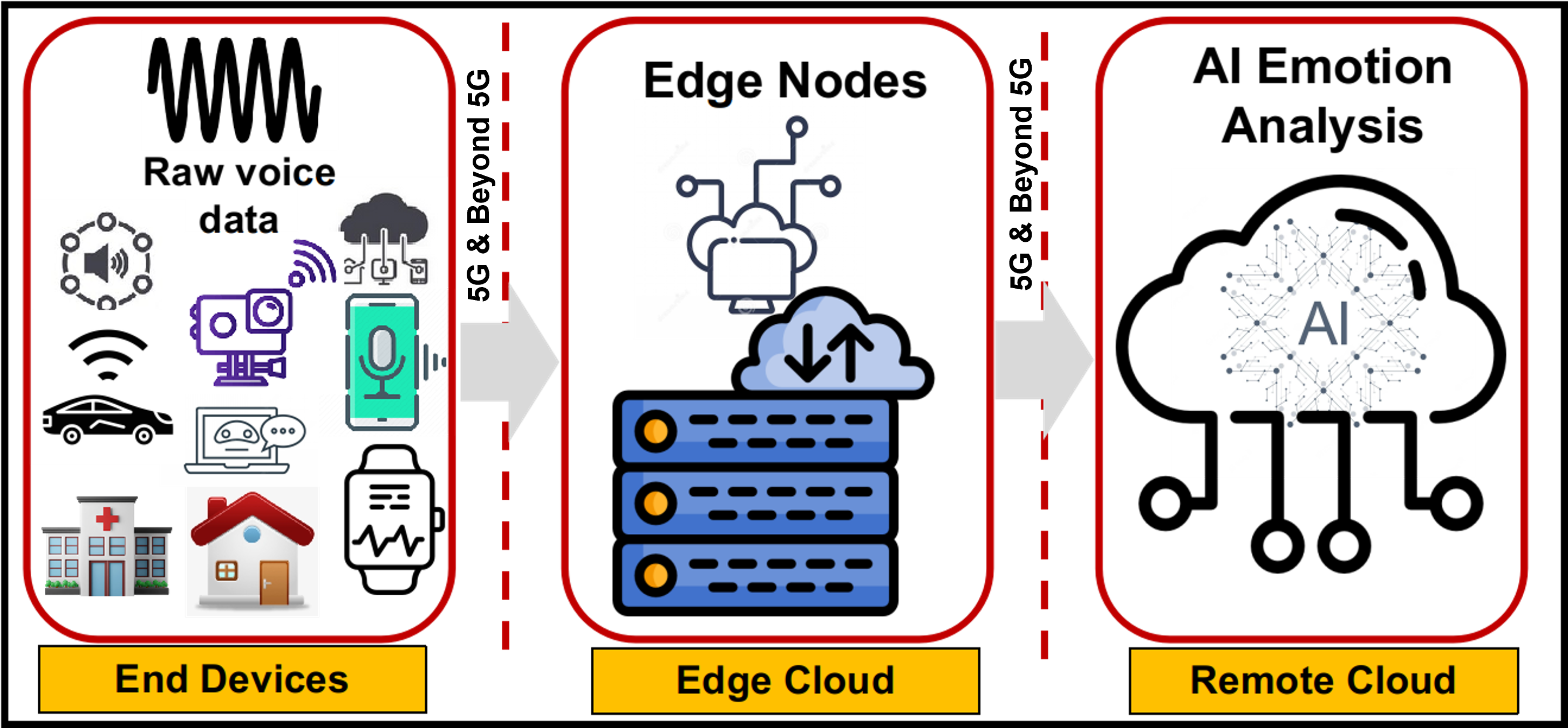}
\caption{State-of-the-art non-speech emotion-sensing system that transmits raw speech signals over the communication network for emotion analysis.}
\label{fig:state}
\end{figure*}

Emotion recognition systems gained traction and their performance has increased dramatically owing to cutting-edge DL-enabled face, voice, language, and psychological signal models. 
The majority of emotion-sensing services use a system paradigm in which raw data collected via IoT sensors is transferred to a distant server for processing and decision-making as shown in Figure \ref{fig:state}. Since emotion recognition systems are intended to detect and classify emotion in real-time, it is critical to create a system with an acceptable level of end-to-end latency, from data acquisition to emotion classification. 

Existing studies on speech-based emotion detection mainly focus on improving the accuracy of the systems for enabling their real-time applications \cite{latif2020deep}. In these systems, they use audio conversations and pass them to different deep-learning models to predict different emotions \cite{latif2023transformers}. The audio conversations used in these systems contain scripted speech datasets. However, emotions do not always exist in speech. Non-speech signals like screams also contain rich emotions. The timely identification of screams has a wide range of applications in public spaces, healthcare \cite{latif2020speech}, age care, rescue services, crime control, and gaming. The delays in classification might result in a fatality, and the latency issue becomes even more concerning. In addition, state-of-the-art emotion-sensing applications follow the system model in which raw speech is transmitted to the remote server for processing and decision-making. Such systems are successful in real-life, however, they involve complete sharing of speech over the communication network, which may lead to adverse consequences to people's privacy \cite{latif2020federated, manzoor2022federated}. Edge computing addresses the latency concerns and privacy related concerns by processing data at the edge server in a federated environment \cite{latif2022ai}.

Most of the emotion-sensing services follow the system model in which raw speech is transmitted to the remote server for processing and decision-making. This has been shown in Figure \ref{fig:state}. Such systems are successful in real-life, however, they involve complete sharing of speech over the communication network, which may lead to adverse consequences to people's privacy \cite{jaiswal2020privacy, manzoor2022federated}. Speech signal contains sensitive information about the message, speaker, gender, language, etc., which may be misused by eavesdropping adversary without users' consent \cite{latif2020federated}.

In this paper, we propose a framework for IoT-based edge computing-enabled non-speech emotion recognition systems. We have made the following contributions:
\begin{enumerate}
 \item We propose to leverage edge computing to design a low-latency non-speech-based emotion recognition system for resource-constrained devices. 

 \item We develop a computationally efficient non-speech emotion detection system by utilising knowledge distillation.
\item We provide a detailed discussion on the potential of using non-speech emotions for various applications such as healthcare, rescue services, etc. 

\item We show the effectiveness of the proposed framework by evaluating the system using two publicly available datasets. Results show that our proposed model can achieve better performance compared to the well-known MobileNetV3-small model \cite{DBLP:journals/corr/abs-1905-02244} and provide better computational efficiency. 
\end{enumerate}

\section{Applications of Scream Recognition}
\label{Sec:rel}

In this section, we will provide a brief description of the available scream detection systems from the literature. The objective here is to provide a non-exhaustive list of works based on non-speech emotion recognition-based systems.

\subsection{Healthcare and Rescue Services}
Understanding non-verbal emotions is a growing area of investigation in healthcare research. In the recent pandemic, many researchers investigated the prospect of producing an early forecast of COVID-19 by understanding the sound of coughs. Similarly, many elderly patients care researchers sought to comprehend the patients' requirements by detecting and interpreting coughs and screams \cite{imran2020ai4covid, alam2018healthband}. Psychologists are also seeking to detect and comprehend non-verbal emotional activity to identify various psychiatric disorders. We have also seen remarkable growth in assistive technology for the sick and the elderly in recent years. Several of them were in the form of wrist/hand bands and lightweight sensors placed on and within the human body. These assistive technologies discern non-speech-based human emotions and other human capabilities by combining cutting-edge IoT technology with broad-ranging learning algorithms. Alam et al. \cite{alam2018healthband} designed a portable hand band for scream detection for dementia patients. The band is inexpensive and convivial (that is, HealthBand) for monitoring the activities of dementia patients and the vigilance of the people in all the trouble.

Another application of non-verbal speech detection which is partly related to healthcare is rescue services. Scream detection techniques play a vital role in locating the victim (human or animal) in catastrophes such as earthquakes, wildfires, etc. Since rescue-related operations are time sensitive and require vigilance, using AI/ML techniques for scream detection can aid in the rescue of the trapped victims under debris and in the burning sites. Saeed et al. \cite{saeed2021initial} designed an AI/ML-enabled scream detection system mounted on a small autonomous vehicle that can help rescue victims from a burning site. The scream detection model in this system was based on support vector machines (SVM) and long short-term memory (LSTM). Given the dynamic inherent nature of their occupations, mobile workers are constantly at risk of being hurt; scream detectors can aid in the rescue of personnel in the event of an emergency \cite{saeed2021initial,kalbag2022scream}.  


\subsection{Crime Control}
With the advent of AI and advanced communication technologies, crime detection is becoming a booming research direction. Scream detection has a direct relation with violent crimes and using AI/ML techniques aided by the data gathered from multiple sensors deployed across urban spaces is an interesting application. Laffitte et al. \cite{laffitte2019assessing, laffitte2017automatic} proposed an ML-based screaming/shouting detection mechanism. Marteau et al. \cite{marteau2020audio} proposed deep learning-based methods to identify audio events such as screams, glass breaks, gunshots, and sprays. Unfortunately, the crimes of racism, harassment, and rape are on the rise in society \cite{leon2021presence,borumandnia2020prevalence,walia2021border}. Urban spaces are increasingly becoming unsafe for women, transgenders, and other genders. The application of scream detection with other surveillance technologies can help protect people from these crimes. In 2018, Seoul Metro, Korea installed scream detectors in women's bathrooms in metro stations to ensure women's safety\footnote{Seoul Metro installs scream detection system in women's bathrooms, Korea Herald. \url{https://www.koreaherald.com/view.php?ud=20180323000689}. Access Date: 15 April 2023.}. Similarly, the Paris metro company is also considering AI-enabled scream detection technologies in the subways to prevent abnormal situations \cite{laffitte2016deep, laffitte2019assessing}. We strongly believe that scream detection techniques can help reduce a lot of crimes. 


 \begin{figure*}[!ht]
\centering
\includegraphics[width=.8\textwidth]{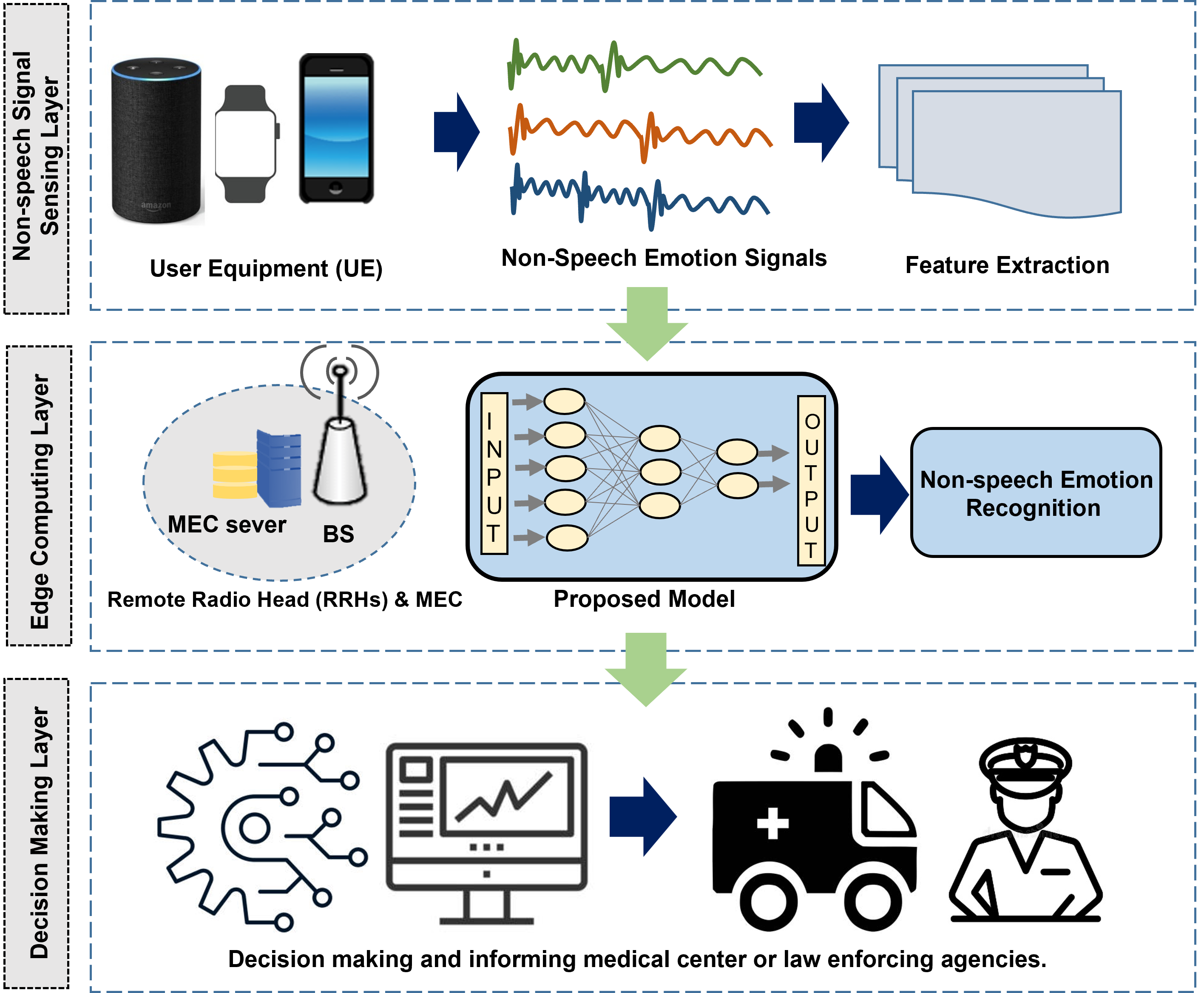}
\caption{Components of the proposed system: (1) a \textit{non-speech signal sensing layer} that collects speech data from mobile phones, personal assistants, or smartwatches, and converts it into speech features; (2) an \textit{edge computing layer} that uses these features to train a deep learning model for non-speech emotion recognition; and (3) a \textit{decision-making layer} that analyses non-speech emotions and makes decisions based on their positivity or negativity, potentially sending alerts to an ambulance or police.}
\label{fig:proposed}
\end{figure*}

\subsection{Home Applications}
Scream detection is becoming an essential tool accompanying visual monitoring in homes, security applications, nursing homes, etc. For instance, Huang et al. \cite{huang2010scream} proposed an energy continuity-based approach for feature extraction from at-home audio recordings and then used the support vector machine (SVM) for identifying the screams in the recorded data. O'Donovan et al. \cite{o2020detecting} proposed and evaluated an ML-based method for scream detection (behavioral disorders) in publicly available datasets of at-home voice recordings. They used a pre-trained CNN for learning scream detection from an audio dataset. For validation, they chose the dataset from the famous TV show ``Supernanny" because of its similarity with the clinical data. These results indicate that using public datasets for learning the behavioural disorders (screams and tantrums) and then using them for clinical recordings is more appropriate than collecting a corpus of expensive private sensitive clinical data for training the behavioural disorder detector models. 




Domestic violence and violent relationships have increased dramatically in recent years \cite{usta2021covid}. Scream detection techniques, along with IoT-enabled voice assistive technology, can identify these heinous crimes and potentially save many individuals from harm. Fleury et al. \cite{fleury2008sound} used recordings from eight microphones placed in a flat and use speech recognition algorithms to determine various elements of human speech. This notion of autonomous voice detection may be expanded to identify screaming and assist a large number of individuals suffering from domestic violence and abusive relationships \cite{chen2017home}. Despite its technical feasibility,  significant ethical and privacy issues remain in creating these security and surveillance applications.

\subsection{Gaming Applications}
Screams are a significant component of speech, and comprehending the emotions associated with these screams is vital for speech detection and translation systems. Conventional speech dialogue datasets do not contain enough screams for learning and investigating the screams properly. Mori et al. \cite{mori2020gaming} used combat games to record a dialogue corpus with more samples of screams and used that corpus for analysing the nature of social screams. Virtual and augmented reality-based games are getting popular for training rescue workers, first responders, ordinary people, and kids for dealing with emergencies. The scream detection system is expected to play a vital role in the gamified preparation for dealing with emergencies \cite{lochmannova2023using}.

\section{Proposed System}

In this section, we discuss the details of the proposed edge-based non-speech emotion recognition system. The proposed framework in this study (Figure \ref{fig:proposed}) offers data collection and analytics support within the 5G network architecture, which is commonly referred to as the network data analytics framework (NWDA) according to 3GPP standards. Basically, the proposed framework can invoke NWDA functions to provide these two core functionalities; first, data collection from network functions (NFs) i.e., local data processing, and second, data analytics and non-speech emotion recognition.  Specifically, the proposed framework serves the following key layers including sensing, edge computing, and decision-making layers. 


\subsection{Non-Speech signal Sensing Layer}
In order to perform edge-based non-speech emotion recognition, we devised a specialised speech sensing layer. The non-Speech signal-sensing layer enables end-user devices to collect non-speech data from cyber-physical space. Nowadays, we witness a vast proliferation of smart edge devices ranging from smartphones, smartwatches and personal assistants such as Alexa Home, Google Assistant, Siri, etc. These edge devices, specifically the personal voice assistants take speech input and perform certain actions such as turning on lights, shutting down appliances, or playing demanded music. In most of these applications, the input speech is converted into text that helps to determine the user's intent using natural language processing and the specific action that needs to be taken. In our proposed framework, the non-Speech signal sensing layer inputs raw speech signals and essential pre-processing is performed to mitigate the background noise effects. The layer processes the input signal and converts it into the features, i.e., melspectrograms. As shown in Fig \ref{fig:proposed}, in the non-Speech signal sensing layer, the audio features are propagated to the network instead of transmitting raw signals directly to the cloud server.





\subsection{Edge Computing Layer}
In our proposed  model, the system has edge and core layers. The edge layer consists of end devices, i.e., mobile phones, tablets, etc., and the edge server which is placed near the base station (BS) as mobile-edge computing (MEC) server \cite{liu2020toward}. It is assumed that the MEC server can process edge signals and is also able to perform analytics on non-speech emotion data \cite{muhammad2021emotion}. It is also assumed that the edge server has enough computational resources to execute data-intensive machine-learning tasks. The proposed framework adopts the 3rd generation partnership project (3GPP) release 16 support for machine learning-based data-driven optimization \cite{3gpp-23.791}. 

  

The edge server is an important component of our proposed architecture. The edge server leverages low-consumption computational and storage hardware such i.e., edge cloudlets \cite{taleb2017multi} and operates within a radio access network (RAN) in the close vicinity of end-users \cite{liu2020toward}. In our proposed architecture (as shown in Figure \ref{fig:proposed}),  the edge server not only performs traffic aggregation gateway and network service control, but it also acts as an intelligent edge server that is responsible for the identification of screams from the given speech features. 




Deploying deep learning models on edge computing devices is an active area of research and many techniques have been proposed to improve the latency and performance of these models. Some prominent techniques include pruning, quantization, knowledge distillation, and training computationally efficient models. In our experiments, we use knowledge distillation \cite{hinton2015distilling} to train a small and efficient model that can be deployed on edge devices.  

\vspace{2mm}
\subsubsection{Knowledge Distillation}
\begin{figure}[!h]
\centering
\includegraphics[width=.5\textwidth]{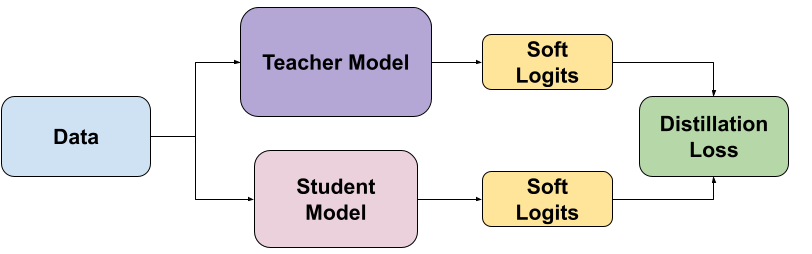}
\caption{Response-based knowledge distillation. The output logits from the student and teacher model are used to calculate the distillation loss between the student and teacher.}
\label{kd}
\end{figure}

The process of knowledge distillation as the name suggests is the method of transferring knowledge from a larger computationally expensive model to a relatively smaller model. The larger and smaller models are called the teacher and student models respectively. Thus, knowledge distillation consists of three principal components: (1) knowledge; (2) distillation algorithm; and (3) teacher-student architecture. While there are now multiple methods of distillation algorithms we selected the response-based algorithm. As shown in Figure \ref{kd}, the hypothesis is that the student model will learn to mimic the predictions of the teacher model. This can be achieved by using a loss function, termed the distillation loss, that captures the difference between the logits of the student and the teacher model respectively.

As this loss minimizes overtraining, the student model will improve at making the same predictions as the teacher. In the offline training scheme, the teacher model is first trained and the weights are then frozen. Next, we train the student model using the distillation loss and the logits from the teacher model as targets. Following is the equation of the distillation loss.

\begin{multline*}
L_{d} = \alpha T^2 \cdot KL \Big(\operatorname{softmax}(T^{-1} \cdot f(T,x)), \\
\operatorname{softmax}(T^{-1} \cdot g(T,x))\Big),
\end{multline*}

where:

$L_{d}$: the loss function for knowledge distillation\\
$\alpha$: a hyperparameter that controls the trade-off between the classification loss and the distillation loss\\
$T$: the temperature hyperparameter used to soften the logits (outputs of the last layer before softmax) of the teacher and student models\\
$KL$: the Kullback-Leibler divergence, a measure of how different two probability distributions are\\
$\operatorname{softmax}$: a function that converts the logits to probabilities\\
$f(T,x)$: the logits of the teacher model for input $x$\\
$g(T,x)$: the logits of the student model for input $x$\\

\vspace{2mm}
\subsubsection{Teacher Model}
\begin{figure}[!ht]
\centering
\includegraphics[width=.4\textwidth]{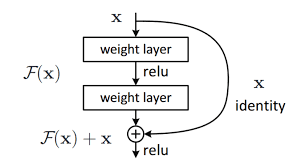}
\caption{Residual layer. The weights of the input are added to the outputs from proceeding convolution layers.}
\label{res}
\end{figure}

\begin{figure}[!ht]
\centering
\includegraphics[height=12cm, width=.21\textwidth]{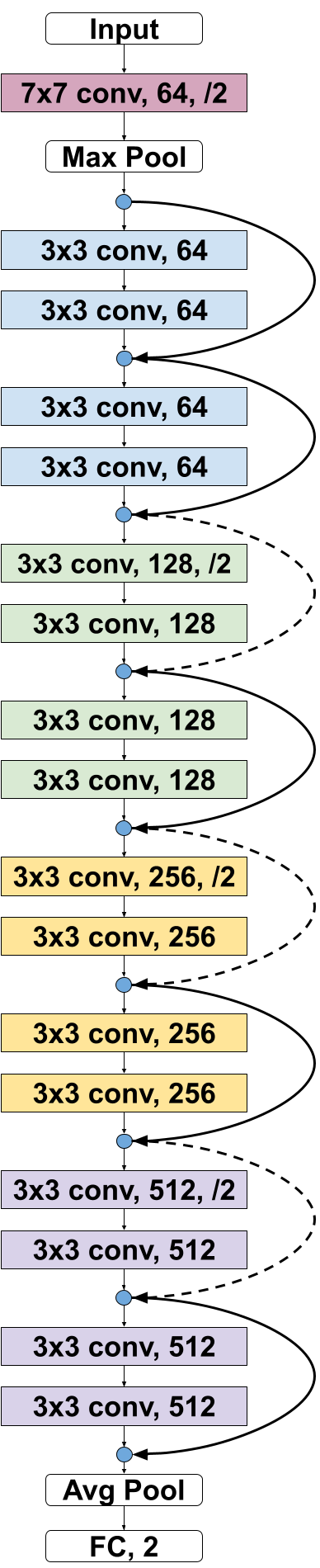}
\caption{Resnet18/Teacher architecture. The coloured blocks represent the convolution layers and the respective kernel sizes, output filters and the reduction of input size. The lines between the convolution layers represent residual connections, whereas the dashed represents that the input of the residual goes through a convolution for dimension consistency. The final FC layer is a dense layer.}
\label{resnet18}
\end{figure}

Generally, for the teacher model a larger and deeper network is chosen so that it performs well on the task at hand. We chose ResNet18 \cite{DBLP:journals/corr/HeZRS15} as our teacher model. ResNet18 contains 18 residual blocks stacked together which alleviates the degradation and vanishing gradient problem. Figure \ref{res} shows a single layer, where the outputs of the previous layer are added to the outputs of the next layer, and Figure \ref{resnet18} depicts the complete architecture of ResNet18 used for the teacher model. To ensure consistent size prior to addition, the input may undergo an operation that aligns it with the output dimensions. This operation is typically a convolution.\\

\vspace{2mm}
\subsubsection{Student Model}
\begin{figure}[!ht]
\centering
\includegraphics[width=.5\textwidth]{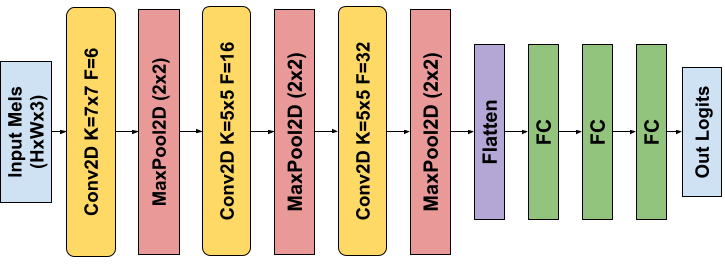}
\caption{Student model. Each coloured block is a convolution layer where $K$ denotes the kernel size and $F$ denotes the number of output filters. The final FC layers are dense layers.}
\label{student}
\end{figure}

Unlike the teacher model, the student model is smaller and shallower, making it more computationally efficient. Our proposed student network simply consists of 3 convolutional layers followed by 3 fully connected layers. Figure \ref{student} provides details on the relatively shallower student model. The first convolution layer has a kernel size of $7\times7$ and the remaining two have $5\times5$ each. The number of filters in each layer is 6, 16, and 32, and after each convolution layer, we apply a maxpool layer of $2\times2$ window size. The fully connected layers have the outputs in this order: 128, 64, 2. For regularisation, we add a dropout after each convolution and fully connected layer with a dropout probability of 10\%.  The non-linear activation chosen between the layers is Rectified Linear Unit or commonly referred to as ReLU.

\subsection{Decision Making Layer}
We train our scream detection model at edge devices. Edge device communicates with the cloud  server via cellular infrastructure and shares model outcome. The cloud server is responsible for scream analytics, decision-making, and storage services. We deploy the proposed classifier to the edge devices to perform the identification tasks. The output from the model is sent to the decision-making layer that can take necessary action based on the situation.  For instance, if scream emotions are classified as negative emotions i.e., the person is in pain or sorrow, the cloud system will send an alert to the healthcare centre or to the police because the person can be injured or hurt by someone. If the scream emotions are classified as positive emotions such as joyous screams, the cloud system would not be sending any alerts.

\section{Experimental Setup}
\label{experi}
In this section, we preset the details on datasets, input features, and training protocol. 

\subsection{Datasets Used in Our Experiments}


\begin{itemize}
    
\item \textit{ASVP-ESD:} The Audio, Speech, and Vision Processing Lab Emotional  Sound database (ASVP-ESD) is a dataset that contains speech and non-speech utterances. There are a total of 12625 audio samples, that are collected from various sources. The samples include both male and female speakers and the emotions are boredom (sigh, yawn), neutral, happiness (laugh, gaggle), sadness (cry), anger, fear (scream, panic), surprise (amazed, gasp), disgust (contempt), excite (triumph, elation), pleasure (desire), pain (groan), disappointment. 

\item \textit{VIVAE:} The Variably Intense Vocalizations of Affect and Emotion Corpus (VIVAE) dataset \cite{holz_natalie_2020_4066235} consists of human non-speech emotion utterances. The full-set contains a total of 1085 samples from eleven speakers. The utterances are divided into three positive (achievement/ triumph, sexual pleasure, and surprise) and three negative (anger, fear, physical pain) emotional states. The audio has a sampling rate of 44.1kHz.

\item \textit{DEMAND:} We use this dataset to evaluate the performance of the proposed framework in noisy conditions. The Diverse Environments Multichannel Acoustic Noise Database (DEMAND) dataset \cite{thiemann_joachim_2013_1227121} provides recordings that can be used to evaluate algorithms using realistic noises captured in various real-world settings. The dataset spans over 6 categories. 4 of these are in an inside environment and the remaining category samples are collected in an outdoor setting. The dataset recordings are available in 48kHz and 16kHz sample rates. The audio was initially recorded for a long duration and afterwards trimmed to a total of 300 seconds each.

\end{itemize}

\subsection{Input Features}
In speech and audio research, melspectrograms are a popular method to represent input signal \cite{latif2021survey}. Similarly, we chose to represent our audio samples as melspectrograms using a short-time Fourier Transform of size 1024, a hop size of 256, and a window size of 1024. The frequency range was chosen between 0-8kHz and a total of 128 bands were computed. Each melspectrogram was normalized in the range of $[-1,1]$. Since the sample utterances were not consistent in length, we decided on a cutoff of 3 seconds for larger audios and padded the smaller ones with zeros, giving us consistent 3-second audios. Before converting the audios to melspectrograms of higher sampling rates we resample them to 16kHz. This sampling rate is kept consistent throughout the experiments and datasets.

\subsection{Training Protocol}

The training of the classification tasks was done using an Nvidia GeForce RTX 3090 24-GB GPU with PyTorch as the framework of choice. The model was trained on a batch size of 64 using the binary cross entropy loss as the criterion. The weights of each layer were randomly initialized with Adam as the optimizer with the following parameters: $\beta_1=0.9$, $\beta_2=0.9999$, $\epsilon=1e^{-8}$. We experimented with multiple learning rates and found that  a learning rate of $1e^{-5}$ gave better results with less training time. 

All experiments were conducted on 80\% and 20\% random splits for training and testing respectively. For our scream detection task conducted on the ASVP-ESD dataset, we had to balance the scream and non-scream utterances as scream utterances totalled 1170 samples. To balance the dataset, we randomly selected 1170 non-scream utterances, giving us effectively 2340 samples to train the scream detector. 

During experimentation, we noticed that the model would overfit, resulting in a high train and low test accuracy. This high bias could be attributed to small dataset sizes and to mitigate this problem of high bias we added augmentations to our training data. These augmentations were composed of stretching and contracting audio samples and adding a low-amplitude Gaussian noise. Additionally, we randomly masked the time and frequency axes of the computed melspectrograms. This augmentation scheme proved helpful in terms of model generalisability and training.

\section{Results and Discussions}
\label{resu}
The objective of this proposed system is to detect non-speech emotions in the communication network using edge computing. We primarily focus on two types of experiments: scream detection and scream emotion detection. The former separates scream utterances from non-scream ones and the latter classifies whether a person's scream is in a situation of danger or duress. This is motivated by the fact that not all screams warrant an investigation or point to emergency situations. It is thus important to cluster screams in positive and negative categories so that the user can be notified of the type of scream detected by the system. For each of our experiments, we provide a comparison between the teacher, student, and a MobileNetV3-small model \cite{DBLP:journals/corr/abs-1905-02244}, a popular choice for edge computing. For brevity, we refer to MobileNetV3-small simply as MobileNetV3s in the proceeding sections.

\begin{figure*}[!h]
\centering
\includegraphics[width=.90\linewidth]{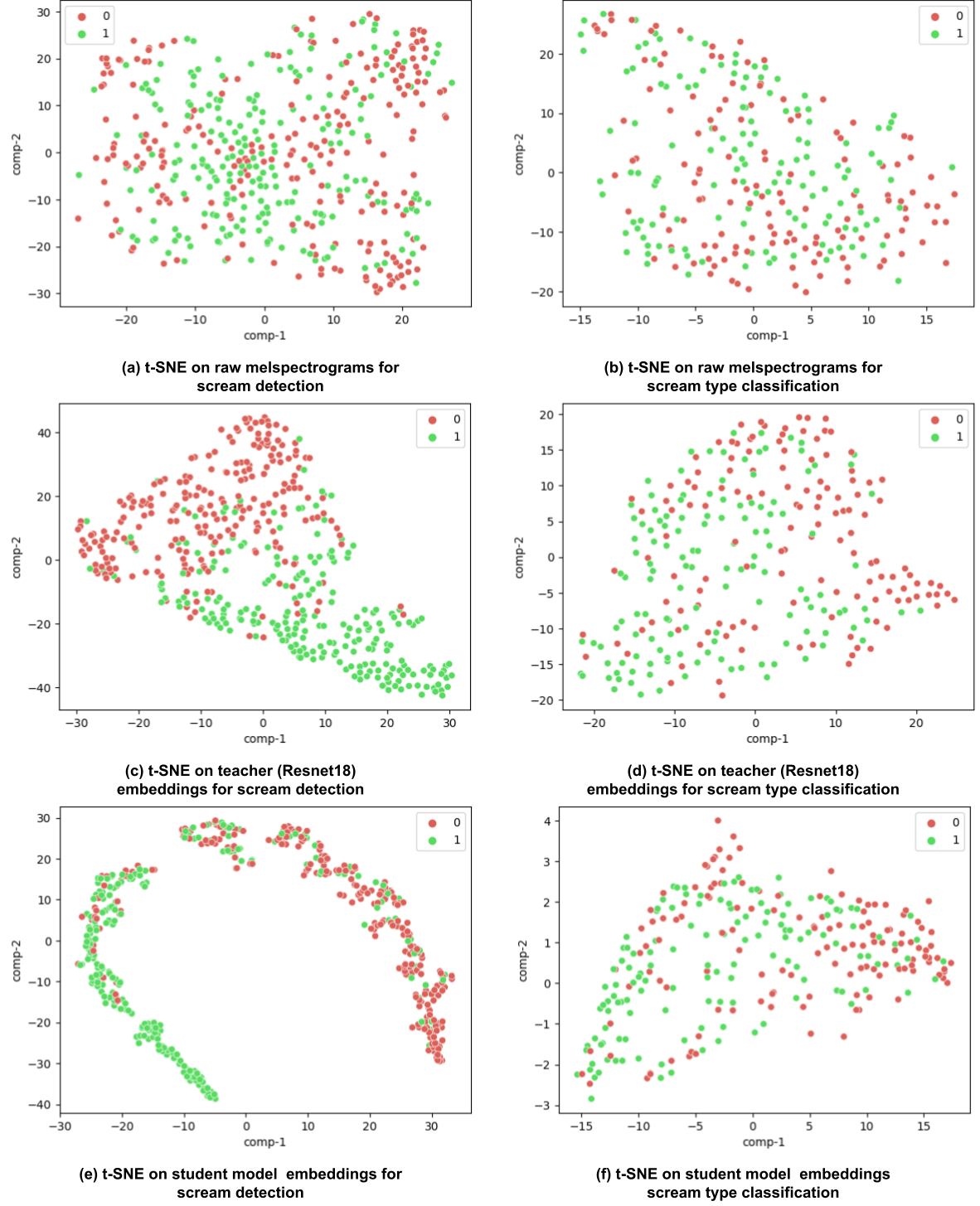}
\caption{t-SNE plots on raw melspectrograms and the penultimate activations (embeddings) of the teacher and student model. There is no clustering on raw melspectrogreams, however, the t-SNE plots after using the trained models show samples getting clustered.}
\label{fig:tsne}
\end{figure*}

\subsection{Non-Speech Emotion Detection}
In this section, we present the results of our scream detection and scream type classification tasks. The former was conducted on the ASVP-ESD dataset and the latter on the VIVAE dataset.  The results of the experiments are presented in Table \ref{exp1}. We can observe that the model performs well in classifying scream and non-scream samples. However, the model struggles to cluster the utterances into positive and negative categories. 


\begin{table}[]
\centering
\scriptsize
\caption{Classification results for experiments without added noise.}
\begin{tabular}{|c|c|c|}
\hline
\textbf{Model}             & \textbf{Task}              & \textbf{Accuracy} \\ \hline
\multirow{2}{*}{Teacher}   & Scream Detection           & 85.47             \\ \cline{2-3} 
  & Scream Type Classification & 73.80             \\ \hline

\multirow{2}{*}{MobileNetV3s} & Scream Detection           & 76.92             \\ \cline{2-3} 
  & Scream Type Classification & 65.80             \\ \hline
  \multirow{2}{*}{Student (proposed)}   & Scream Detection           & \textbf{80.58}             \\ \cline{2-3} 
  & Scream Type Classification & \textbf{67.16}             \\ \hline
\end{tabular}
\label{exp1}
\end{table}

We can see how the samples might be clustered by using the t-SNE algorithm \cite{JMLR:v9:vandermaaten08a} for dimensionality reduction. In Figure \ref{fig:tsne} we provide T-distributed stochastic neighbourhood embedding (tSNE) plots using the raw melspectrograms from the datasets and the penultimate activations of the teacher and student model. The plots illustrate that there is little to no clustering within the raw melspectrograms. This changes when we train the model on scream classification tasks. We can observe that for the scream detection task, we get distinct clusters for both the teacher and student models. The clustering is sparse for the scream type classification from the teacher and student models but still better than the results from the raw melspectrograms. This sparsity might explain the complexity of this task.

\subsection{Evaluations in Noisy Conditions}

In real-world scenarios, the background is often not static in nature which causes the inclusion of noise in the environment. To simulate a real-world scenario, we added noise from the DEMAND dataset to our audio samples and tested our evaluations in a simulated noisy real-world environment. The noise samples were randomly selected from the following environments: bus, metro, cafe, kitchen, and office. Within each noise sample, we randomly select a chunk equal to the input audio sample. Table \ref{exp2}. summarises the results of this experiment. To test the robustness of the model performance we added noise only in the test split. The results show that the model generalises well when the noise is added for evaluation only.

\begin{table}[!ht]
\scriptsize
\centering
\caption{Classification results for experiments in a noisy setting.}
\begin{tabular}{|c|c|c|}
\hline
\textbf{Model}             & \textbf{Task}              & \textbf{Accuracy} \\ \hline
\multirow{2}{*}{Teacher}   & Scream Detection           & 82.26             \\ \cline{2-3} 
  & Scream Type Classification & 69.74             \\ \hline

\multirow{2}{*}{MobileNetV3s} & Scream Detection           & 76.92             \\ \cline{2-3} 
  & Scream Type Classification & 60.15 
  \\ \hline

  \multirow{2}{*}{Student (proposed)}   & Scream Detection  & \textbf{78.60}             \\ \cline{2-3} 
  & Scream Type Classification & \textbf{66.42}             \\ \hline
\end{tabular}
\label{exp2}
\end{table}

To further test the performance of the student model in noisy situations, we compare the evaluation results for each noise category used previously. Table \ref{exp3}. highlights that the results for individual noises do not decrease drastically, giving us overall similar results as discussed earlier. 

\begin{table}[!h]
\caption{Classification results of student model for individual noise categories.}
\centering
\begin{tabular}{|c|c|c|}
\hline
\multirow{2}{*}{\textbf{Task}}              & \multirow{2}{*}{\textbf{Noise Type}} & \multirow{2}{*}{\textbf{Accuracy}} \\
  &    &  \\ \hline
\multirow{4}{*}{\begin{tabular}[c]{@{}c@{}}Scream\\ Detection\end{tabular}}           & Cafe & 72.86 \\ \cline{2-3} 
  & Kitchen & 78.85 \\ \cline{2-3} 
  & Office  & 80.13 \\ \cline{2-3} 
  & Metro& 73.50  \\ \hline
\multirow{4}{*}{\begin{tabular}[c]{@{}c@{}}Scream Type\\ Classification\end{tabular}} & Cafe & 65.68 \\ \cline{2-3} 
  & Kitchen & 63.10 \\ \cline{2-3} 
  & Office  & 63.84 \\ \cline{2-3} 
  & Metro& 65.31 \\ \hline
\end{tabular}
\label{exp3}
\end{table}

\subsection{Computational Complexity}
\begin{figure}[!h]
\centering
\includegraphics[width=1\linewidth]{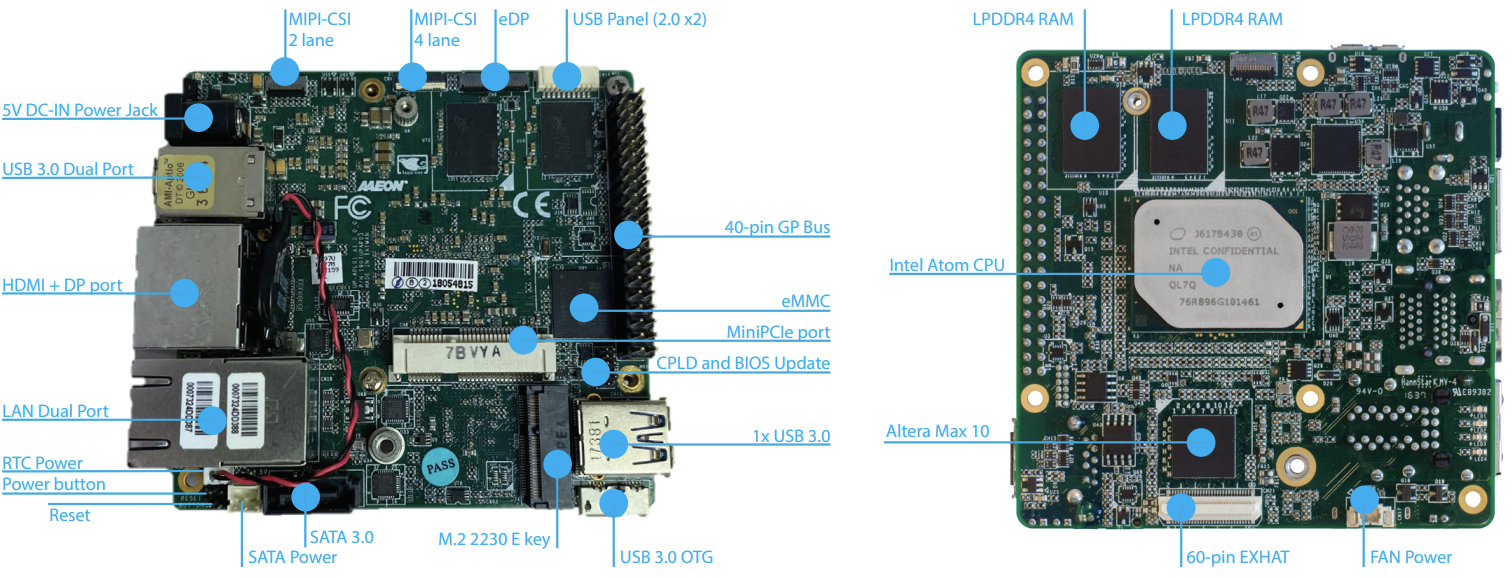}
\caption{Labelled diagram of UpBoard with Intel(R) Atom E3940 CPU, Source \cite{upboard}.}
\label{fig:up}
\end{figure}

 The benchmarks presented in this section are conducted on an UpBoard (shown in Figure \ref{fig:up}) having an Intel(R) Atom E3940 CPU operating at 1.60GHz, with an onboard memory of 4GBs. The tests are recorded on the standard Ubuntu 20.04 LTS operating system (we did not use the server/headless version) and PyTorch version 1.13. The results are concluded without the use of post-processing features offered by PyTorch. These features claim to lower the inference time but generally with some trade-offs. 

\begin{figure}[!h]
\centering
\includegraphics[width=\linewidth]{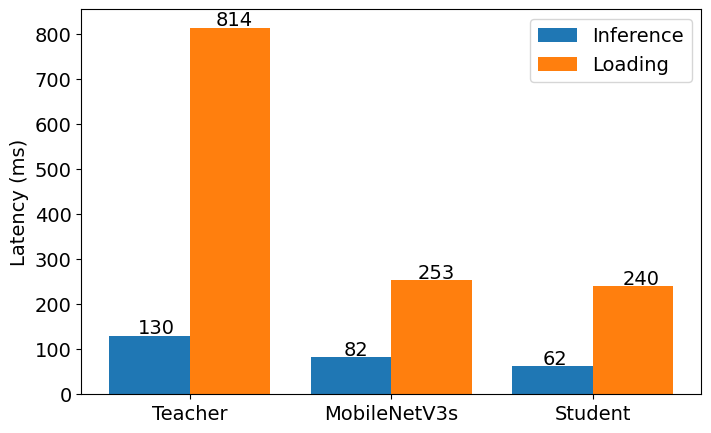}
\caption{Comparison of latency and loading times in milliseconds between teacher, student (proposed), and MobileNetV3s models. }
\label{fig:bench}
\end{figure}

 The total time to load the model and the time it takes for a single forward pass of an utterance are two metrics that are of primary concern when it comes to deploying models in production. Likewise, we benchmark these metrics for the teacher, student, and MobileNetV3s models. Figure \ref{fig:bench} provides a graphical comparison between the benchmarks of the mentioned models, which shows that in terms of latency, the student model is almost twice as fast as the teacher model and is about 20ms faster than the MobileNetV3s model. Furthermore, there is a significant difference in terms of load times where the student model takes the least time to load into the memory. 
 The difference between these metrics might better be explained by the total parameters and the size of each model presented in Table \ref{params}. A larger number of parameters in a model increases the computational cost. Similarly, a large memory footprint contributes to higher model loading times. 

\begin{table}[!h]
\centering
\caption{Size and total parameters of the teacher, student, and MobileNetV3s models.}
\begin{tabular}{|c|r|r|}
\hline
\textbf{Model} & \multicolumn{1}{c|}{\textbf{\begin{tabular}[c]{@{}c@{}}Total \\ Parameters\end{tabular}}} & \multicolumn{1}{c|}{\textbf{\begin{tabular}[c]{@{}c@{}}Size\\ (MBs)\end{tabular}}} \\ \hline
Teacher        & 11177616      & 42.676 \\ \hline
MobileNetV3s    & 1519906       & 5.844  \\ \hline
Student (proposed)        & \textbf{712778}& \textbf{2.719}  \\ \hline
\end{tabular}
\label{params}
\end{table}

These results highlight the deployment of the proposed system into memory and computation constraint devices such as personal assistants (Alexa Home, Google Assistant, and Siri). Personal assistants are then made an effective choice for the tasks of scream detection and classification. Based on scream positive or negative nature, these personal assistants can send alerts to the Police or medical centre to help the person.

\section{Conclusions} 
\label{con}
This paper presents a knowledge distillation-based non-speech emotion identification system for edge computing. We covered various applications of non-speech emotion identification and provided a case study based on real-life scenarios. We evaluated system performance based on two publicly available datasets. We designed our experiment setup to distinguish scream sound from other utterances and classify non-speech utterances based on their emotional states. To highlight the robustness of the proposed system, we also evaluated these experiments by adding typical real-world background noises to our inputs to mimic real-world scenarios. Results demonstrated that the proposed framework provides better computational efficiency compared to the well-known MobileNetV3 and achieves improved performance. These results showed the feasibility and effectiveness of our proposed non-speech emotion identification system in communication networks. In the future, we aim to study the energy efficiency of the proposed non-speech emotion identification system.

\end{document}